\documentclass[twocolumn,secnumarabic,amssymb,nobibnotes,aps,prd,longbibliography]{revtex4-2}
\usepackage{graphicx}
\usepackage{epstopdf, epsfig}
\usepackage{epsfig}
\usepackage{graphicx}
\usepackage{amsmath}
\usepackage{amssymb}
\usepackage{multirow}
\usepackage{bbold}
\usepackage[table,xcdraw]{xcolor}
\usepackage{subfigure}
\usepackage{threeparttable}
\usepackage{array}
\usepackage{wasysym}
\usepackage{textcomp}
\usepackage{color}
\usepackage{hyperref}

\begin{document}

\title{Quantized orbital-chasing liquid metal heterodimers directed by an integrated pilot-wave field}%
\thanks{A calibration issue with the acceleration measurement in our system led to acceleration amplitudes ten times higher than the actual values. In this version, correct acceleration values have been updated. A correction note has also been published alongside the journal version [\href{https://link.aps.org/doi/10.1103/PhysRevFluids.5.053603}{Phys. Rev. Fluids 5, 053603 (2020)}].}

\author{Jianbo Tang}%
\email[Email:]{jianbo.tang@unsw.edu.au}
\affiliation{Department of Biomedical Engineering, School of Medicine, Tsinghua University, Beijing 100084, China}
\affiliation{Current address: University of New South Wales, Sydney, NSW 2052, Australia}

\author{Xi Zhao}%
\thanks{The first two authors contributed equally}
\affiliation{Key Laboratory of Cryogenics, Technical Institute of Physics and Chemistry, Chinese Academy of Sciences, Beijing 100190, China}
\affiliation{School of Future Technology, University of Chinese Academy of Sciences, Beijing 100049, China}

\author{Jing Liu}%
\email[Email:]{jliubme@tsinghua.edu.cn}
\affiliation{Department of Biomedical Engineering, School of Medicine, Tsinghua University, Beijing 100084, China}
\affiliation{Key Laboratory of Cryogenics, Technical Institute of Physics and Chemistry, Chinese Academy of Sciences, Beijing 100190, China}
\affiliation{School of Future Technology, University of Chinese Academy of Sciences, Beijing 100049, China}


\begin{abstract}
\vspace{4mm}
A millimetric bouncing droplet sustained on a vibrating bath becomes a moving wave source (particle) through periodically interacting with the local wave field it generates during the droplet-bath impact. By virtue of such particle-wave duality, the macroscopic hydrodynamic system imitates enigmatic behaviors of the quantum realm. Here we show that it is possible to create an integrated pilot-wave field to better prescribe the droplet trajectories, via amplified bath capillarity. This is demonstrated with a liquid metal droplet-bath system in which the local wave field generated by droplet bouncing is superposed by the global wave field induced by bath meniscus oscillation. The resulting dual pilot-wave configuration enables a class of directional chasing motions of two bound dissimilar droplets (heterodimers) in multilevel hydrodynamic traps (orbits), featuring two quantized regime parameters, namely the interdroplet binding level and the orbit level. We investigate the dynamics of the vibrating liquid metal bath, with its level-split ring wave field and its peculiar vortex field being highlighted. We also rationalize the exotic droplet motions by considering the interdroplet particle-wave interactions mediated by the integrated pilot-wave field. It is revealed that a temporal bouncing phase shift between the two droplets in the heterodimers, due to size mismatch, gives rise to their horizontal propulsion, while their spatial binding regime exclusively determines the collective chasing direction. It is further evidenced that the horizontal in-orbit chasing motion is directly related to vertical droplet bouncing. Our findings unveil the integrated pilot-wave field as a trail towards improved droplet guiding, thereby extending the hydrodynamic particle-wave analogy to optical systems and beyond.
\end{abstract}

\maketitle

\section{Introduction}
\label{sec 1:Intro}
A hydrodynamic pilot-wave system features millimetric bouncing droplets self-propelling on a vibrating fluid bath. The droplet behaviors emerging from the system strongly depend on the dynamics of both the bath and the droplets \cite[]{protiere2006particle, bush2015pilot}. The study of the former, which dates back to Michael Faraday \cite[]{faraday1831xvii}, has been focusing on the nonlinearity and threshold instability of the periodically forced bath \cite[]{douady1990experimental}. This classic problem has been greatly enriched by the introducing of discrete wave sources (bouncing droplets of the same fluid) to the system by Yves Couder, Emmanuel Fort and co-workers \cite[]{couder2005bouncing, couder2005dynamical}. It has been shown in their pioneering works that, below the Faraday threshold $\gamma_\textit{F}$, a vertically bouncing droplet can be long-term sustained and also be carefully tuned to \textit{walk} horizontally through a particle-wave association, that is, the droplet interacts with the waves it generates on the bath. The creation of particle-wave duality in a macroscopic hydrodynamic system has profound influences and meaningful implications since wave-mediated motions are frequently encountered in various physical systems, from the classical interpretation of the electron-phonon wave interactions of Cooper pairs in the Bardeen–Cooper–Schrieffer theory of superconductivity \cite[]{rohlf1994modern} to biological locomotions on water surface \cite[]{Roh24446} and to astronomical scale planetary/stellar motions in the gravitational wave field \cite[]{Flanagan_2005}, which themselves might not be readily accessible. Considerable efforts have been made ever since to explore both individual \cite[]{couder2006single, dorbolo2008resonant, eddi2009unpredictable, fort2010path, harris2013wavelike, harris2014droplets, andersen2015double, filoux2017walking, saenz2018statistical, lieber2007self, protiere2008exotic} and collective droplet behaviors \cite[]{eddi2008wave, eddi2009archimedean, eddi2012level, filoux2015strings, galeano2018ratcheting} in the droplet-bath system, in particular for their quantum analogies \cite[]{couder2006single, eddi2009unpredictable, harris2013wavelike, andersen2015double, saenz2018statistical}.

Despite the diversity of previously observed droplet behaviors, they mostly share the same particle-wave association regime in which the horizontal droplet motions are solely piloted by the droplet-generated local wave field, either by the droplets themselves or by their partners. To a large extent, such a single pilot-wave configuration restricts the search for informative droplet motions to a narrow acceleration range close to but below $\gamma_\textit{F}$, where the walking conditions are satisfied \cite[]{protiere2006particle, molacek2013drops}. One exception is the ratcheting droplet pairs reported by \textcite[]{eddi2008wave} and \textcite[]{galeano2018ratcheting}, a configuration that enables switchable horizontal motions of dissimilar droplet pairs at accelerations well below $\gamma_\textit{F}$.

It has been demonstrated that tuning bath dynamics favors better prescribing droplet trajectories. However, it is typically realized through either auxiliary bath modifications (e.g., engineered bath structures \cite[]{filoux2017walking, saenz2018statistical, filoux2015strings, Sungar2017PhysRevFluids, Sungar2018Chaos, Tambasco2018Chaos}) or bath movement control (e.g., a rotating bath frame \cite[]{fort2010path, harris2014droplets}). In particular, \textcite[]{Sungar2017PhysRevFluids, Sungar2018Chaos} has demonstrated that protruding pillars above the vibrating bath surface can lead to a hydrodynamic effect reminiscent of the optical Talbot effect. To enrich the content of the pilot-wave hydrodynamics as well as to extend its implications, strategies that enable new droplet motion regimes and at the same time avoid previous formalities will be highly desirable. To this end, an attempt is made here to achieve improved droplet guiding without compromising the simplicity and elegance of this classical system.

The superposition nature of waves implies that foreign wave fields other than the droplet-generated local wave field can be introduced to provide further confinements for the droplets. One mechanism to produce such an additional wave field in the droplet-bath system could be dynamic meniscus, since a vibrating fluid bath is always attended by an oscillating meniscus at its boundary region due to capillarity \cite[]{de2013capillarity}. Generally, the oscillating meniscus is considered as a troublesome feature which previous studies tried to circumvent, since it brings both experimentation difficulties and theoretical assessment complexities to the system \cite[]{douady1990experimental, henderson1994surface}. By contrast, here we significantly magnify the capillary effect by selecting a liquid metal droplet-bath system. We show that the previously believed problematic oscillating meniscus can produce a global pilot-wave field, which is indeed advantageous for droplet guiding. Directed by the dual pilot-wave field, dissimilar droplet pairs in our system are able to exhibit well-defined collective orbital motions with double quantized states.

In one-to-one correspondence, we show that the hydrodynamic particle-wave association directed by an integrated pilot-wave field is reminiscent of the wave-matter interaction in optical systems, and the chasing droplet pairs become counterparts of optical \textit{heterodimers}$-$two optically bound particles of unequal sizes \cite[]{yifat2018reactive, sukhov2015actio}. It has been demonstrated that a perpendicular incident light (electromagnetic waves) can trigger transverse motions of optical heterodimers, since the size mismatch between the two particles leads to their asymmetric interactions (scattering) in the presence of the external electromagnetic wave field. When a second electromagnetic field (a circular optical trap) is further applied to create a global confinement, the collective motions of the heterodimers are guided along the optical trap, resulting in optically directed orbital chasing motion \cite[]{yifat2018reactive}. In the optical system, the spatial interparticle distance and the motion trajectory of the heterodimers are modulated by the interparticle electrodynamic interaction (optical binding) and the predefined global electromagnetic wave field (optical trap), respectively \cite[]{yifat2018reactive}. We show that directed by an integrated hydrodynamic pilot-wave field, the particle-wave association between the droplets and the local wave field generated by their partners provides the interdroplet binding, whereas the global wave field acts as a multilevel hydrodynamic trap which leads to orbital motion. Strikingly, despite the fact that the optical heterodimers and the hydrodynamic heterodimers interact with wave fields of different natures, their motion regimes (e.g., configuration and direction) are, however, surprisingly alike. In addition, the directional transverse motions in both systems stem from the symmetry breakdown of their respective particle-wave association, which itself is a result of the breakdown of size symmetry. Therefore, it is attempting that the theoretical frameworks that describe the two systems, which feature different physical interactions and distinct scales, may share a similar form.  Encouragingly, the results reported here demonstrate the possibility to extend the hydrodynamic pilot-wave analogy to other wave-mediated systems as well as the feasibility to guide droplet motions with an integrated pilot-wave field.

\begin{figure*}
\centering
\includegraphics[width=1\linewidth]{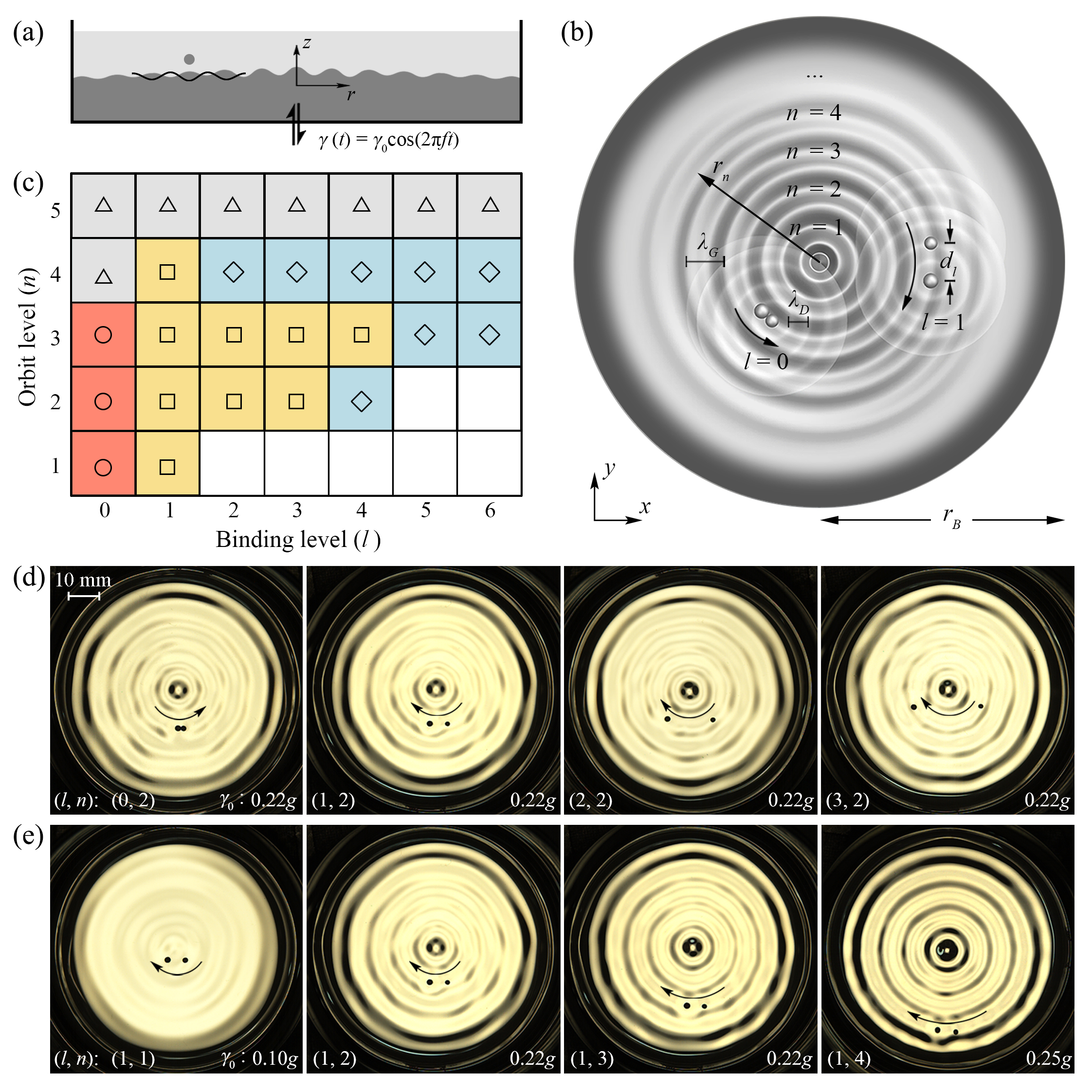}
\label{FIG1}
\caption{\label{figure1} (a) Schematic side view of the liquid metal droplet-bath system. The bath is filled with a bottom liquid metal layer and a top alkaline electrolyte layer. The vibrating bath and the bouncing droplet induce a global wave field $\mathcal{W}^G$ and a local wave field $\mathcal{W}^D$, respectively. (b) Schematic top view of the directional orbital-chasing motion of the liquid metal heterodimers and the definition of the regime parameters. ${\lambda{}}_G$ and ${\lambda{}}_D$ represent the wavelength of $\mathcal{W}^G$ and $\mathcal{W}^D$, respectively. (c) The regime diagram of the orbital-chasing motion. ($\Circle$) The large droplet $\mathcal{D}_L$ chases the small droplet $\mathcal{D}_S$; ($\Box$) $\mathcal{D}_S$ chases $\mathcal{D}_L$; ($\Diamond$) $\mathcal{D}_S$ chases $\mathcal{D}_L$ but the heterodimers either cease or derail before a full-circle orbiting is completed (partial orbiting regimes); ($\triangle$) High-orbit levels in which the confinement becomes too weak to sustain the heterodimers. Regimes requiring $\textit{$d_{l}$}>2\textit{$r_{n}$}$ are geometrically invalid and left blank. (d) Top-view snapshots of the orbiting heterodimers with different binding levels \textit{l} (\textit{n} = 2 fixed). (e) Top-view snapshots of the orbiting heterodimers in different orbit levels \textit{n} (\textit{l} = 1 fixed). The arrows indicate the direction of orbiting and chasing.
}
\end{figure*}

\section{System configuration}
\label{sec 2: sysconfig}
The system configuration used in this study is sketched in Fig. 1(a). A cylindrical glass container (inner radius $\textit{r}_B$ = 38 mm) is fixed on an electromagnetic shaker (not shown) to form the vibration platform \cite[]{zhao2018electrically}. The container is filled with a liquid metal layer of thickness \textit{h}$_1$ = 10.0 $\pm{}$ 0.5 mm. The liquid metal used here is a eutectic alloy of gallium and indium (EGaIn, 75.5\% Ga and 24.5\% In by weight percentage) which has the lowest melting point of the Ga-In binary system $T_m\sim$15 \textcelsius, density ${\rho{}}_1$ = 6280 kg m$^{-3}$ and viscosity ${\nu{}}_1=5.13\times{}{10}^{-7}$ m$^2$ s$^{-1}$. The liquid metal is covered by an alkaline electrolyte top layer (0.5 mol L$^{-1}$ NaOH aqueous solution, \textit{h}$_2$ = 10.0 $\pm{}$ 0.5 mm, ${\rho{}}_2=1012$ kg m$^{-3}$, ${\nu{}}_2=9.32\times{}{10}^{-7}$ m$^2$ s$^{-1}$) to prevent the oxidation of the liquid metal surface so that it behaves as a simple fluid. Consequently, the droplet bouncing `trampoline' in the current case is a liquid metal-electrolyte interface with an exceedingly high interfacial tension \textit{$\sigma{}$} (0.456 N m$^{-1}$ measured with the pendant drop method). Before each experiment, the platform is checked and adjusted carefully with a level gauge to minimize off-axis vibration.

During the experiment, the bath is forced vertically with a sinusoidal acceleration $\gamma{}\left(t\right)={\gamma{}}_0\cos{\left(2\pi{}ft\right)}$, where ${\gamma{}}_0$ is the peak driving acceleration, $f$ the driving frequency (fixed at 40 Hz), and \textit{t} denotes time. ${\gamma{}}_0$ can be continuously adjusted from zero to well beyond the Faraday threshold ${\gamma{}}_F$ = 0.32\textsl{\textrm{g}}, (the liquid metal-electrolyte interface) of the current system with an accuracy of $\pm{}$0.005\textsl{\textrm{g}}, where \textsl{\textrm{g}} is the gravitational acceleration. Note that the Faraday threshold ${\gamma{}}_F$ may be altered under the influence of the meniscus waves in our system. In the current study, ${\gamma{}}_F$ is measured as the threshold value when the meniscus wave pattern breaks. In addition, the Faraday threshold of the upper electrolyte/air interface is found to be higher than that of the liquid metal/electrolyte interface located below. Therefore, the Faraday waves of the electrolyte/air interface do not emerge within the acceleration range in which the droplet motions are investigated. A plastic pipette is used to generate millimetric liquid metal droplets which can be long-term sustained and readily manipulated on the bath. The extrusion volume of the pipette is adjusted to control the size of droplets and the actual droplet size is measured by post-image analysis. The uncertainty of droplet size measurement is within 5\%. To place the droplets on the liquid metal-electrolyte interface, the tip of the pipette is inserted near the interface to avoid any air entrainment. The vertical/horizontal droplet motion and the evolution of the bath wave field are captured by a high-speed camera with recording frequency up to 3200 Hz. For the vertical-view recordings, a LED light source is illuminated from the opposite side of the bath to the camera, while for the horizontal-view recordings, the light source is illuminated from above the bath, atop the camera.

\section{Horizontal droplet motion}
\subsection{The orbital-chasing motion and regime diagram}
\label{3.1_the orbital motion and regime diagram}
Provided that suitable droplet size and forcing conditions are satisfied, two dissimilar liquid metal droplets autonomously form a locked pair (heterodimer) and start revolving collectively yet stably around the bath center, following the annular rings of the global wave field $\mathcal{W}^G$ (Figs. 1b$-$1e and Video 1). As will be detailed later, the so-called simple bouncing mode is excited in the current system, where the droplets impact the bath once per bath vibration \cite[]{protiere2006particle}. Noticeably, the chasing liquid metal heterodimers take discrete interdroplet binding distance \textit{d$_{l}$} (Fig. 1d) and orbit radius \textit{r$_{n}$}$_{ }$(Fig. 1e), where \textit{l} and \textit{n} represent the binding level and the orbit level, respectively. As presented in the regime diagram in Fig. 1c, modulating the variable set (\textit{l}, \textit{n}) produces highly diversified orbital motion regimes of the heterodimers. See Video 2 of the Supplemental Material for a demonstration of different regimes.

Among all the regimes, we observe two types of interdroplet binding which lead to chasing motions in opposite directions. Note that the chasing direction here is defined by the size of the leading droplet of a heterodimer rather than the direction of its angular velocity, as the orbiting can always be changed from clockwise to anticlockwise or \textit{vice versa} by switching droplet position. In the short-range binding regime ($l=0$, $\Circle$ in Fig. 1c), the heterodimers advance with the large droplet ($\mathcal{D}_L$) chasing behind the small droplet ($\mathcal{D}_S$), while in the long-range binding regimes ($l\geq{}1$, $\Box$ and $\Diamond$ in Fig. 1c), $\mathcal{D}_L$ takes the lead ($\mathcal{D}_S$ becomes the follower) and thus the chasing reverses. The motion of the long-range binding liquid metal heterodimers is particularly similar to that of the optical heterodimers \cite[]{yifat2018reactive}, but the achievable regimes in the hydrodynamic system appear to be more diversified than its optical counterpart. It should be pointed out that two dissimilar droplets placed in adjacent orbits can also form a special type of long-range binding heterodimers (mostly realized when \textit{l} = 1 and 2). However, this kind of cross-orbit motion will not be detailed in the current study.

We note that the center of the global orbital motion of the liquid metal heterodimers is located at the bath center rather than along the two-droplet alinement. The latter is the case for the orbiting walker pairs as demonstrated in Ref. \cite{protiere2006particle}, \cite{couder2005dynamical} and \cite{protiere2008exotic}. We shall also point out that, different from the ratcheting droplet pairs \cite{eddi2008wave, galeano2018ratcheting}, the chasing direction of the heterodimers is exclusively determined by the binding regime \textit{l}, and it is not observed to reverse as we progressively sweep ${\gamma{}}_0$ from the bouncing threshold ${\gamma{}}_B$ all the way to the Faraday threshold ${\gamma{}}_F$. These distinct features indicate the emergence of new particle-wave association mechanisms in the current system.

\subsection{The parametrics and quantization of the orbital motion}
\label{3.2_the orbiting parametrics and quantization of the orbital motion}
We combine high-speed imaging and post visual tracking  to analyze the horizontal orbital motion of individual droplets to shed light on their collective behaviors. The motions of the heterodimers are captured from above the bath at the same frequency as the driving acceleration ($\textit{f}=40$ Hz). In doing so, no out-of-plane motion of the bath will be seen, although it is constantly vibrating, thereby improving the tracking accuracy. Using the method described in Ref. \cite{danelljan2014accurate} and \cite{gao2017deep}, the typical error for horizontal motion tracking can be kept less than $1/10$ of the characteristic droplet diameter. Examples of the horizontal droplet motion tracking are presented in Video 3 of the Supplemental Material. The orbiting trajectories of individual droplets in the $x-y$ plane in different orbit levels \textit{n} (\textit{l} = l fixed) are plotted against \textit{t} in Fig. 2(a), which shows well-defined trajectories and long-term stability of the orbital motions. Further analyses reveal a definitive feature of the collective droplet motions in the current system that the orbital motion is accompanied by cyclically repeated instantaneous variations. For instance, calculating the horizontal orbiting velocity of the droplets shows that, while the magnitude of their cycle-averaged velocity $\bar{v}$ remains constant, that of their instant velocity $v$ changes constantly yet periodically (Fig. 2b). Interestingly, when comparing Fig. 2(a) and Fig. 2(b), it is evident that $v$ changes precisely at the same frequency (i.e., the same number of cycles) as the orbiting motion and even the fluctuations are well repeated for all cases. Also, we note that there exist noticeable differences in the velocity as well as that the velocity changes between $\mathcal{D}_S$ and $\mathcal{D}_L$.

\begin{figure*}
\centering
\includegraphics[width=1\linewidth]{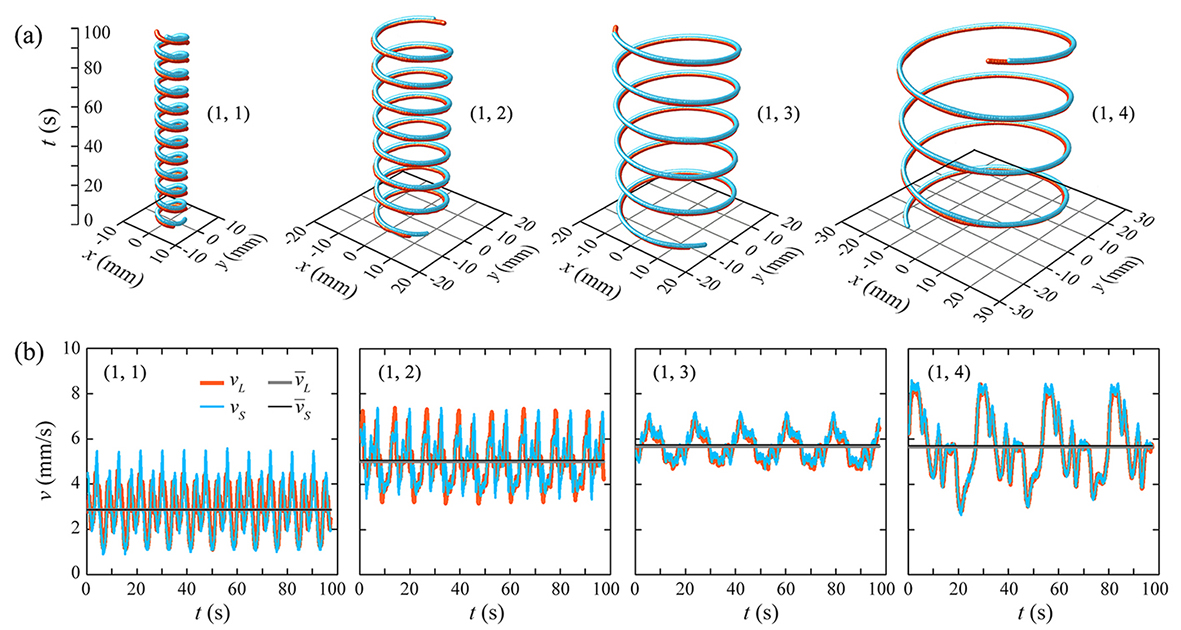}
\label{FIG2}
\caption{
(a) The orbiting trajectories of the large droplet $\mathcal{D}_L\ $(red) and the small droplet $\mathcal{D}_S$ (blue) featuring different orbit levels \textit{n} (\textit{l }= 1 fixed). (b) The magnitudes of the instant velocity $v$ and the average velocity $\bar{v}$ of the heterodimers in (a) as a function of time. The errors in the measurement of \textit{v} and $\bar{v}$ are $\textless$ 0.2 mm s$^{-1}$ and $\textless$ 0.04 mm s$^{-1}$, respectively.
}
\end{figure*}

As shown in Fig. 3(a), time-dependent $v$ directly results in instantaneous variations of $d_l$, even when the two droplets are close to each other ($\textit{l}=0$, small binding distance). Again, two locked droplets always experience changes in $d_l$, but the cycle-averaged value ${\bar{d}}_l$ remains constant. It is confirmed that $d_l$ also changes at the orbiting frequency of each heterodimer. In addition, we find that ${\bar{d}}_l$ is exclusively determined by the binding level \textit{l} and it is independent of both the orbit level \textit{n} and the driving acceleration ${\gamma{}}_0$ (Fig. 3b). This evidence suggests that the orbital-chasing motion of the heterodimers in the current system is affected locally in the orbits. As will be discussed in the next section, the perturbations the heterodimers experience during their orbital motion indeed originate from a peculiar vortex structure formed on the vibrating liquid metal bath. Nevertheless, as shown in Fig. 3(c), the discrete values of ${\bar{d}}_l$ and $r_n$ both fall onto a linear curve, leading to the quantization of the orbital chasing motions:

\begin{equation}
\label{eqdl}
\begin{aligned} 
\bar{d}_{l} &=\left(l-\varepsilon_{l}\right) k_{l}, & l &=1,2,3, \ldots \\
\end{aligned}
\end{equation}
\begin{equation}
\label{eqrn}
\begin{aligned} 
r_{n} &=\left(n-\varepsilon_{n}\right) k_{n}, & & n=1,2,3, \ldots 
\end{aligned}
\end{equation}
One exception is the short-range binding regime $l=0$ where ${\bar{d}}_0\sim\left(d_L+d_S\right)/2$ with $d_L$ and $d_S$ being the diameters of $\mathcal{D}_L$ and $\mathcal{D}_S$, respectively. The physical implications of the slopes ($k_l=6.6$ mm, $k_n=6.5$ mm) and offsets (${\epsilon{}}_l=0.2$, ${\epsilon{}}_n=0.4$) of the fitting curves of Eq.(\ref{eqdl}) and Eq.(\ref{eqrn}) can be related to the wave parameters of different pilot-wave fields, a point that will be addressed later.

\begin{figure*}
\centering
\includegraphics[width=1\linewidth]{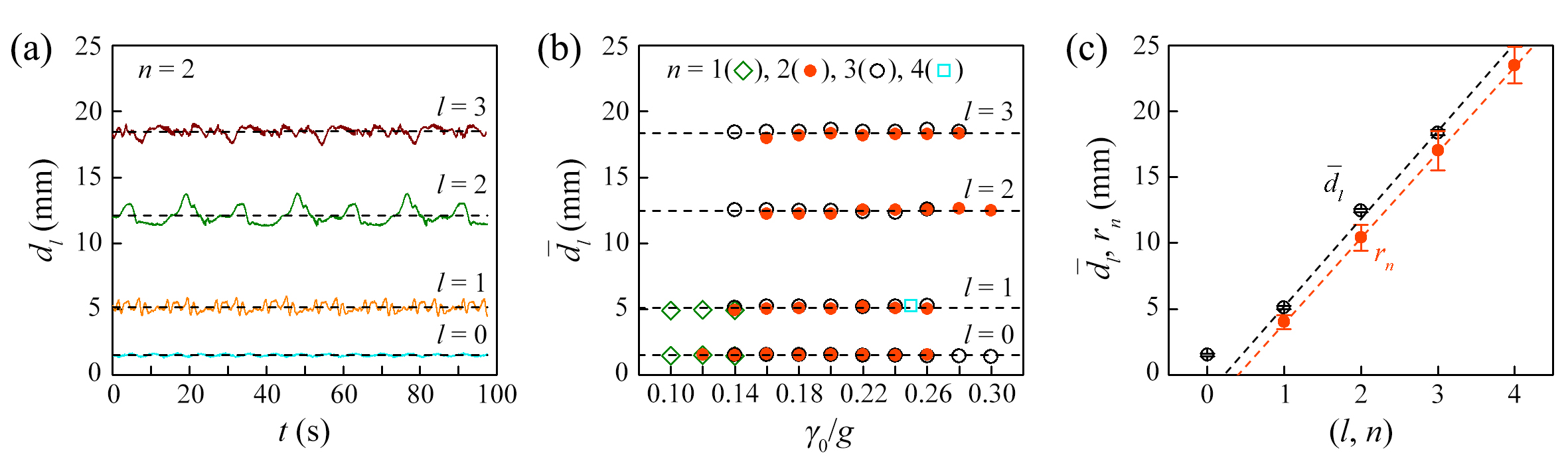}
\label{FIG3}
\caption{
(a) Binding distance \textit{d}$_{l}$ variations of the
heterodimers at different binding levels \textit{l} (\textit{n} = 2 fixed).
(b) Cycle-averaged binding distance ${\bar{d}}_l$ in different orbiting
regimes as a function of the normalized driving acceleration ${\gamma{}}_0$/\textsl{\textrm{g}}.
(c) Quantization of the orbital motion of the liquid metal heterodimers.
In all figures, the solid lines and scatters represent experimental data and the dashed lines are their linear fittings.}
\end{figure*}

\section{Bath dynamics}
\subsection{The profile of the bath wave field}
\label{4.1_the profile of the bath wave field}
One important observation from the orbital motions of the liquid metal heterodimers is that their trajectories are always confined in the concentric rings (orbits) formed on the bath (Figs. 1 and 2). Such circular confinements are provided by a global annular wave field $\mathcal{W}^G$, the formation of which is revealed at the bath vibrating onset. As shown in Fig. 4(a), the bath side views (slightly oblique to show the wave field) captured at its vibration onset show unambiguously that the annular ring patterns are formed by the traveling wave fronts emitted from the bath boarder (see Video 4 of the Supplemental Material). When a single droplet is deposited on the vibrating bath, the local wave field $\mathcal{W}^D$ generated by the droplet impact superposes to $\mathcal{W}^G$, resulting in the formation of the dual pilot-wave field (Fig. 4b). We note, however, that no orbital motion is observed in the single droplet scenarios (see Video 5 of the Supplemental Material).

The wave fronts constructing $\mathcal{W}^G$ are indeed generated by a significant meniscus of the liquid metal bath oscillating at the driving frequency \textit{f} (harmonic)\cite[]{douady1990experimental}. A quantitative assessment of the scale of the meniscus-influenced region as well as the surface profile of both the quiescent bath and the vibrating bath is further performed using a customized micro positioning stage. A sharp-tip stainless-steel probe, which will not be wetted by the liquid metal, is used to detect its surface (liquid metal-electrolyte interface). Since the liquid metal surface is highly reflective, small distortions on its surface upon contact become readily distinguishable, which is recognized as the detecting of the surface during the measurement. With the assistance of a micro positioning stage, both the vertical ($\textit{z}$) and radial ($\textit{r}$) positions of the probe tip are determined with micrometer resolution. The bath vibrating amplitude $\textit{$\Delta{}z$}$ in Fig. 4(c) is obtained by subtracting the measured bath vertical profile under vibration by its quiescent profile (Fig. 4d). Note that the current method only picks up the vibrating maxima of the bath surface and therefore the profile curves in Fig. 4(c) represent the vibrating amplitudes (maximum displacement in the $\textit{z}$ direction) along the radial direction over a whole vibrating period rather than the instantaneous profiles of the bath. The measured vibrating amplitude indicates the total energy of the waves since on the curves the kinetic energy is fully converted into potential energy. 

\begin{figure*}
\centering
\includegraphics[width=0.65\linewidth]{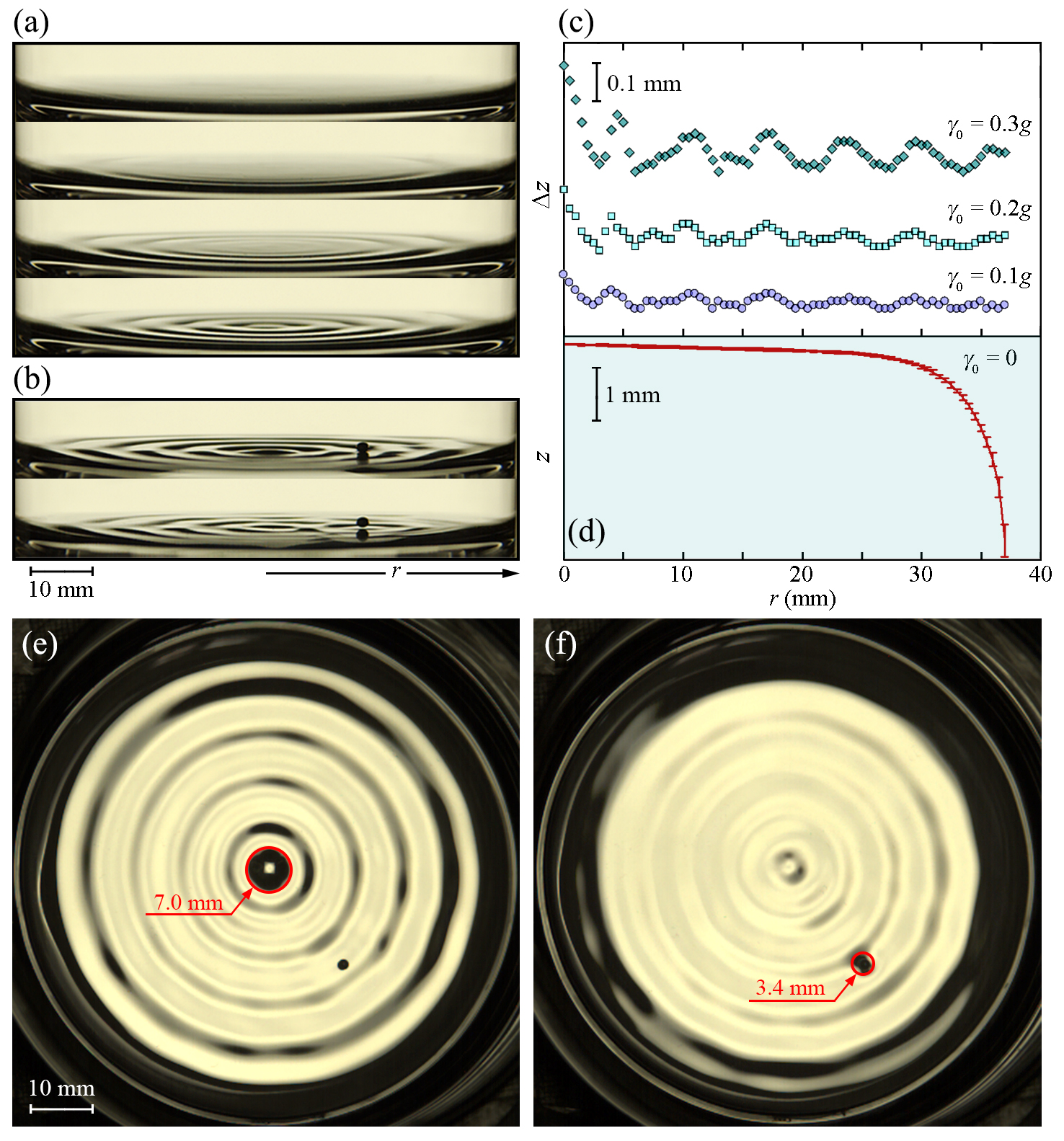}
\label{FIG4}
\caption{
(a) Time-lapse side view snapshots (slightly oblique) of the liquid metal bath at the vibrating onset. From top to bottom, \textit{t} = 0.00 ms, 22.23 ms, 41.67 ms, and 86.12 ms. (b) Side view snapshots of a single droplet bouncing on the vibrating bath (with no horizontal motion). The superposition of the droplet-induced local wave field $\mathcal{W}^D$ and the global wave field $\mathcal{W}^G$ can be seen from the bottom panel. (c) Bath vibrating amplitude ${\Delta{}}z$ measured along the radial direction \textit{r} at different driving accelerations as indicated. (d) Surface profile of the quiescent liquid metal surface. (e), (f) Top-view snapshots featuring the maximum bath deformation at the bath center due to meniscus oscillating (e) and the maximum localized bath deformation induced by droplet impact (f). The maximum size of the dark region (deformed liquid metal surface) at the bath center (e) and the droplet impact site (f) is indicated. ${\gamma{}_0}$ = 0.22\textsl{\textrm{g}}.
}
\end{figure*}

As shown in Fig. 4(d), a giant ($\sim$10 mm) meniscus-influenced region can be seen from the bath profile. The formation of such an anomalous meniscus can be explained by Zisman's rule \cite[]{de2013capillarity}: The high-energy liquid metal surface in contact with the relatively low-energy surface of the glass container adopts a nonwetting regime, a configuration resulting in a large contact angle (large meniscus), despite that the capillary length of the liquid metal $l_C=\sqrt{\sigma{}/({\rho{}}_1-{\rho{}}_2)\textsl{\textrm{g}}} \sim3.0$ mm is comparable to that of common fluids (e.g., water $l_C\sim2.7$ mm). The fact that the meniscus which introduces the global wave field is governed by the relative surface energy of the contacting phases implies that the realization of an integrated pilot-wave field is not limited to the liquid metal droplet-bath system demonstrated here. Systems using common working fluids can also reproduce similar effects by choosing non-wetting (super-hydrophobic or oleophobic) materials, or by reducing the size scale (equivalently, the Bond number), for instance, using sessile drop configuration, to amplify capillarity and generate a significant boundary curvature. 

Emitted by the oscillating meniscus, the amplitude of the propagating wave fronts is affected by two competing processes. The wave amplitude is damped by viscous dissipation while amplified due to the shrinking geometry (diameter) of the waveform when propagating towards the bath center. As shown in Fig. 3(c), the measured surface profiles increase radially towards the bath center in a near-periodic fashion, indicating a stronger influence of the latter. This is guaranteed by the insignificant influence of viscous dissipation (relative to inertial and surface energy) in the current system, as can be expected from its marginal Ohnesorge number $Oh={\nu{}}_1\sqrt{({\rho{}}_1-{\rho{}}_2)/(\sigma{}{\lambda{}}_G)}\sim7\times{}{10}^{-4}$,
where ${\lambda{}}_G=6.4\pm{}0.3$ mm is the measured wavelength of $\mathcal{W}^G$. We note that ${\lambda{}}_G$ can also be deduced from the dispersion relation of surface waves interfacing two finite fluid layers \cite[]{kumar_tuckerman_1994, zhao2018electrically}
\begin{equation}
\label{eqdispersion}
\begin{aligned} 
\omega{}^2 &= \frac{\left(\rho{}_1- \rho{}_2\right)gk+\sigma k^3}{\rho{}_1\coth{\left(k h_1\right)}+\rho{}_2\coth{\left(k h_2\right)}}
\end{aligned}
\end{equation}
where $\omega = 2\pi \textit{f}$ is the angular velocity and $\textit{k} = 2\pi /{\lambda{}_G}$ is the wavenumber. Solving Eq. (\ref{eqdispersion}) for the current system gives \textit{k} = 0.96 mm$^{-1}$ and therefore $\lambda{}_G$ = 6.5 mm, which matches well with the measured value. As discussed earlier, the surface profile measurement picks up only the maxima of the bath vibrating amplitude, so the periodic surface profile indicates a standing wave nature of $\mathcal{W}^G$. This argument can be understood given the superposition of the inward-propagating wave fronts by the outward-propagating ones reflected at the bath center. The traveling-to-standing-wave transition can also be inferred from the vibrating onset process of the bath (Video 4 of the Supplemental Material), which shows that the transition is completed after the first few bath vibrations.

We proceed to discuss the vibration modes of the liquid metal bath in a cylindrical coordinate (\textit{r}, $\textit{$\phi$}$), where $\phi$ is the azimuthal angle. Considering a circular bath under vertical vibration and excluding the influence of the dynamic meniscus, one would expect its instantaneous surface profile to be modulated by

\begin{equation}
\label{eqzrtphi}
\begin{aligned} 
\Delta z(r, \phi, t) &= A_{0}\cos\left(\omega t\right)\cos\left(i\phi \right)J_{i}\left(2\pi r_{i,j}/\lambda{}_G\right) 
\end{aligned}
\end{equation}
where \textit{A}$_0$ is the vibration amplitude at \textit{r} = 0, $\textit{J}_i$(\textit{x}) is the order \textit{i} Bessel function of the first kind, \textit{i} and \textit{j} are integers indicating the vibration modes, and $\textit{r}_{i,j}$ = $\mu^{\text{i}}_{\text{j}}$/\textit{$r_B$} with $\mu^{\text{i}}_{\text{j}}$ being the root of the equation \textit{$J_{\text{0}}^{\prime}$}(\textit{x}) = 0 \cite[]{1999fuac.book.....K}. The formation of the axisymmetric global wave field $\mathcal{W}^G$ (concentric rings) requires the vibration to be independent of $\phi$, which leads to \textit{i} = 0. Eq. (\ref{eqzrtphi}) then becomes
\begin{equation}
\label{eqzrt}
\begin{aligned} 
\Delta z(r, t) &= A_{0}\cos\left(\omega t\right)J_{0}\left(2\pi r_{0,j}/\lambda{}_G\right) 
\end{aligned}
\end{equation}
Eq. (\ref{eqzrt}) suggests that the surface profile is a zero order Bessel function $\textit{J}_0$(2$\pi$\textit{r}$_{0,j}$/$\lambda{}_G$). This axial excitation mode (independent of $\phi$) is the predominant mode in our system according to our experimental observation. Eq. (\ref{eqzrt}) also means that the actual orbit radius $\textit{r}_n$ should satisfy $\textit{J}_0$(2$\pi$$\textit{r}_n$/$\lambda{}_G$) = 0. For the sake of simplicity, we use a linear correlation to approximate $\textit{r}_n$, which should still provide sufficient accuracy to describe the current system (Fig. 2c).

The formation of the global wave field $\mathcal{W}^G$ provides an additional confinement which locks the horizontal motion of the liquid metal heterodimers into its circular orbits. This implies that the droplets acquire the centripetal force needed for their orbital motion from their impact with $\mathcal{W}^G$ and the impact should therefore be made at an inclined
inward-facing wave front. The increasing vibrating magnitude of $\mathcal{W}^G$ towards the bath center (Fig. 4c and Eq. \ref{eqzrt}) implies different effective vibrating accelerations among different orbits, which can be recognized as a level splitting effect \cite[]{eddi2012level}. Due to such level splitting, when the bouncing droplets are located in different orbits, both their vertical and horizontal motions are expected to be altered when traveling in different orbits even under the same driving acceleration.

\subsection{The integrated pilot-wave field}
\label{4.2_the integrated pilot-wave field}
Before accessing droplet motion, it is worthwhile to characterize the vibration amplitude of the two components of the integrated pilot-wave field, $\mathcal{W}^G$ and $\mathcal{W}^D$. With the radial profile of $\mathcal{W}^G$ known, we roughly estimate the relative magnitude of the two wave fields by comparing their maximum deformation. Since the liquid metal surface is smooth yet highly reflective, the sufficiently deformed regions on the bath (including the meniscus region at the bath border, the deformed bath center and the droplet impact site) deviate illuminating light away from the recording camera on top, rendering these regions low brightness (Figs. 4e, 4f, and 5a). The size of the dark regions outlined at the bath center and the droplet impact site gives an indication of the vibration amplitude of $\mathcal{W}^G$ and $\mathcal{W}^D$, respectively (note that the scale of the meniscus-influenced region also can be estimated by measuring the width of the dark annular band surrounding the bright region). Provided that both $\mathcal{W}^G$ and $\mathcal{W}^D$ follow the similar profile of the zero-order Bessel function \cite[]{galeano2018ratcheting}, the deformed (dark) region revealed from the top views can be directly related to the vibration amplitude of the two wave fields. Given the small deformation (on the scale of $\sim$ 0.1 mm) in the current case, a linear approximation can be assumed, which suggests that the magnitude of $\mathcal{W}^D$ is of the same order of magnitude but significantly smaller (about half) in reference to that of  $\mathcal{W}^G$.

\subsection{The bath vortex field}
\label{4.3_the bath vortex field}
Using a digital particle-imaging-velocimetry (PIV) technique, we further demonstrate the formation of a stable counter vortex pair at the liquid metal-electrolyte interface (Fig. 5 and Video 6 of the Supplemental Material). The flow at the interface is visualized by adding boron tracer particles (average size 50 $\mu{}$m) to the bath. Boron particles are selected due to their good material compatibility with the liquid metal and the alkaline solution, as well as their desirable density to settle at the interface. The vortex motion is again captured at the driving frequency (40 Hz) to filter out bath motion (Fig. 5b). In our experiments, rotating the container to different azimuthal positions can change the orientation of the vortex pair and increasing the driving acceleration also leads to the increase of the vortex current magnitude. However, the vortex pair is always found to exist even though care is taken to level the bath.
Based on our observations, we propose that the vortex pair emerges as a result of the azimuthal excitation of the system. Apart from driving the radial excitation mode, which is responsible for the formation of $\mathcal{W}^G$ and independent of the azimuth angle $\phi$ (Eq. \ref{eqzrt}), periodic forcing can also give rise to $\phi$$-$dependent azimuthal excitation, when higher order modes, other than the zero-order mode, are excited (Eq. \ref{eqzrtphi}). When \textit{i} $\geq$ 0, the term $\cos$(\textit{i}$\phi$) in Eq. (\ref{eqzrtphi}) breaks the axisymmetric vibration and the vibration begins to show $\phi$ dependence. Since periodic displacement with a gradient along the vibration direction has been shown to induce transverse  flows \cite[]{hocking1987waves, punzmann2014generation}, the azimuthal excitation should be primarily responsible for the formation of the vortex field in the current system. The structure of the visualized vortex field suggests that the vortex pair is formed through the $\textit{J}_1$(2$\pi$\textit{r}$_{1,j}$/$\lambda{}_G$)) mode excitation.

Being aware of the existence of the counter vortex pair, the instantaneous yet periodic variations of $v$ and $d_l$ in Fig. 2(b) and  Fig. 3(a) can be rationalized. Regardless of their binding regimes, the two droplets of the heterodimers are accelerated by one of the vortices, with which they share the same horizontal rotating direction during half of the orbiting cycle, and are decelerated by the counter vortex during the other half cycle. Due to the mirror symmetry of the vortex pair, the two competing processes cancel each other out after each cycle, rendering $\bar{v}$ and ${\bar{d}}_l$ constant. Therefore, although the orbits are geometrically symmetric, the orbital motion of the heterodimers is disturbed locally by the vortex current. It thus can be seen that the level splitting of the orbits and the formation of vortex pair in the current system alter both the interorbit and in-orbit energy landscapes of $\mathcal{W}^G$. 

\begin{figure*}
\centering
\includegraphics[width=1\linewidth]{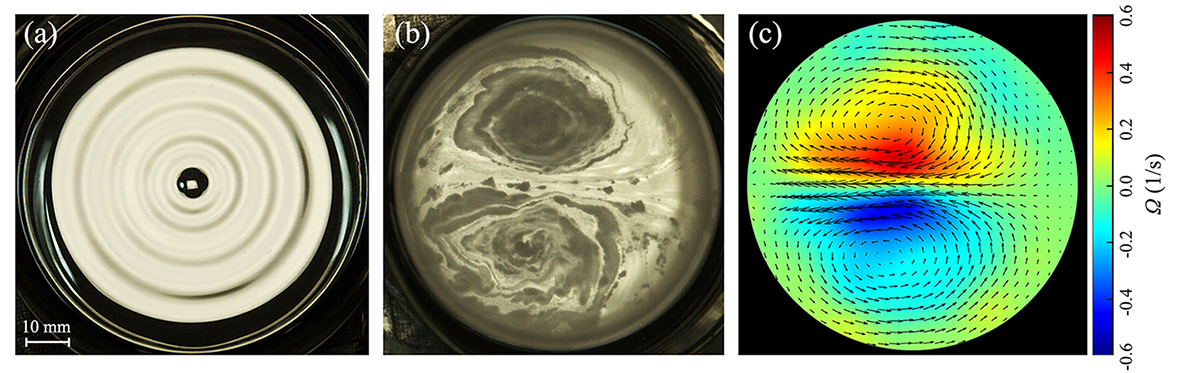}
\label{FIG5}
\caption{
(a) Top view of the vibrating liquid metal bath showing the global annular wave field $\mathcal{W}^G$. (b) Visualized vortex field at the liquid metal-alkaline interface and (c) the corresponding vorticity color map (${\lambda{}}_0=0.16$\textsl{\textrm{g}}).
}
\end{figure*}

\section{Vertical droplet motion}
\subsection{The shifted vertical bouncing}
\label{5.1_the shifted vertical bouncing}
While orbiting horizontally on the bath, the droplets are at the same time bouncing vertically in resonance with the bath vibration (Figs. 6a, 6b), both generating localized harmonic waves. Since a single droplet will never start orbiting until it meets another dissimilar partner, the interactions between individual droplets ($\mathcal{D}_L$ and $\mathcal{D}_S$) and their self-generated local wave fields ($\mathcal{W}_L^D$ and $\mathcal{W}_S^D$), hereafter denoted as $\mathcal{D}_L\leftrightarrow{}\mathcal{W}_L^D$ and $\mathcal{D}_S\leftrightarrow{}\mathcal{W}_S^D$, respectively, should not be responsible for their horizontal motions, nor should their interactions with the meniscus-induced global wave field (${\mathcal{D}_L\leftrightarrow{}\mathcal{W}}^G$ and ${\mathcal{D}_S\leftrightarrow{}\mathcal{W}}^G$). The two-droplet configuration as a prerequisite for their orbital-chasing motion implies that the association between the droplets and the wave fields generated by their partners ($\mathcal{D}_L\leftrightarrow{}\mathcal{W}_S^D$ and $\mathcal{D}_S\leftrightarrow{}\mathcal{W}_L^D$) should provide the horizontal propulsion. Their collective motions further require the horizontal propulsion acting on the two droplets to be in the same direction.

To elucidate the origin of the directional horizontal propulsion, we compare the vertical motion of individual droplets of the heterodimers in different regimes. For vertical droplet motion tracking, the motions are recorded at 1600 Hz to generate 40 data points for each bouncing cycle. The typical spatiotemporal droplet bouncing trajectories for the short-range binding regime (\textit{l} = 0, Fig. 6c) and the long-range binding regime (\textit{l} = 1, Fig. 6d) are reconstructed from the high-speed images \cite[]{protiere2006particle}. The droplet trajectory curves are also generated by the aforementioned visual tracking method (Figs. 6e, 6f, see also Video 7 of the Supplemental Material), which show an excellent match with Figs. 6c and 6d, again indicating good tracking accuracy. Comparing the vertical motion of the two droplets in the heterodimers reveals that, while their vertical bouncing is always synchronized with the bath vibration (i.e., in the simple bouncing regime) \cite[]{protiere2006particle}, their bouncing phase is however different. Independent of their binding level and orbit level, the impact of the large droplet $\mathcal{D}_L$ is always found to be delayed by a time amount of $\Delta{}t$ to that of the small $\mathcal{D}_S$, resulting in a normalized bouncing phase shift $\theta{}/2\pi{}=\Delta{}t/T$ between the two \cite[]{eddi2011information, molacek2013drops}.

\begin{figure*}
\centering
\includegraphics[width=1\linewidth]{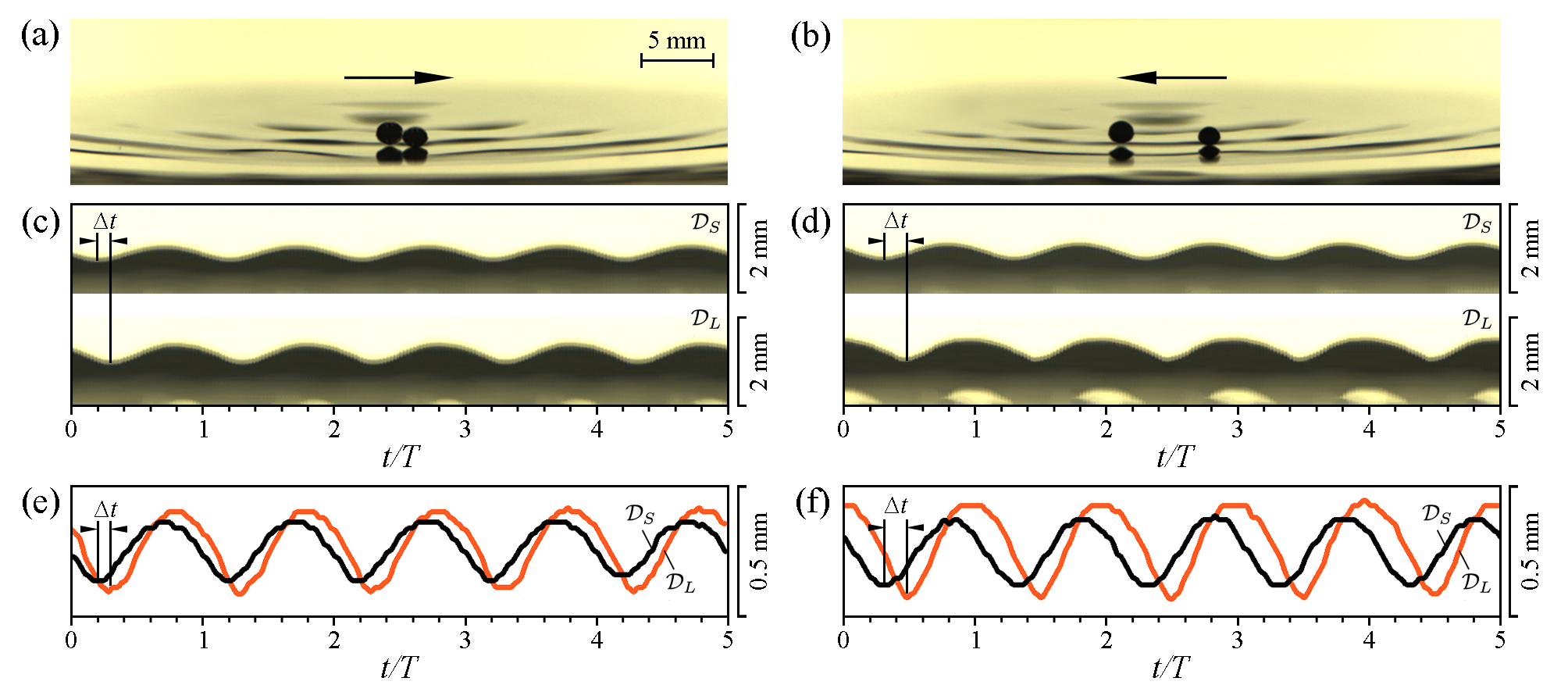}
\label{FIG6}
\caption{(a), (b) Oblique side views of the orbiting heterodimers in the short-range binding regime \textit{l} = 0 (a) and the long-range binding regime \textit{l} = 1 (b). (c), (d) Vertical bouncing trajectories of the small droplet $\mathcal{D}_S$ (upper panel) and the large droplet $\mathcal{D}_L$ (lower panel) as a function of time for five consecutive bouncing periods \textit{T} of a (0, 3) heterodimer (c) and a (1, 3) heterodimer (d). (e) The spatial-temporal trajectories of the same heterodimer in (e) obtained by visual tracking. (f) The spatial-temporal trajectories of the same heterodimer in (d) obtained by visual tracking. The impact time difference $\Delta{}t$ of the two droplets of the heterodimers are indicated.}
\end{figure*}

\subsection{The inter-droplet particle-wave association}
\label{5.2_the inter-droplet particle-wave association}
Being aware of the size-dependent bouncing phase shift, we can rationalize the horizontal propulsion as well as the directionality of the orbital-chasing motions by considering the interdroplet particle-wave association of each heterodimer ($\mathcal{D}_L\leftrightarrow{}\mathcal{W}_S^D$ and $\mathcal{D}_S\leftrightarrow{}\mathcal{W}_L^D$). Under the assumptions that the in-orbit chasing motion is pseudo one dimensional, the droplets and their self-generated waves share the same phase \cite[]{galeano2018ratcheting}, and the Doppler effect is negligible ($v\ll{}{\lambda{}}_Df$), the problem is simplified to the interactions between individual droplets and the waves emitted by their partners with a phase shift $\theta{}$ from a distance $d_l$. If we allow for the shifted evolution of $\mathcal{W}_L^D$ and $\mathcal{W}_S^D$ as presented in Fig. 7(a), the particle-wave association $\mathcal{D}_L\leftrightarrow{}\mathcal{W}_S^D$ and $\mathcal{D}_S\leftrightarrow{}\mathcal{W}_L^D$ of the heterodimers in the \textit{l} = 0 and \textit{l} = 1 binding regimes are depicted in Fig. 7(b)  and Fig. 7(c), respectively. Here $z=0$ represents the zero position of the droplet-generated local wave field $\mathcal{W}^D$.

For both regimes, by the time $\mathcal{D}_L$ lands on the bath, $\mathcal{W}_S^D$ has already evolved for a period $\Delta{}t$ due to the bouncing phase shift. Therefore, the impact site of $\mathcal{D}_L$ has been distorted by $\mathcal{W}_S^D$ upon landing. As shown in Fig. 7(b), when \textit{l }= 0, since ${\bar{d}}_0=\left(d_L+d_S\right)/2<{\lambda{}}_D/3$ in the current system (${\lambda{}}_D=6.5\pm{}0.2$ mm is the measured wavelength of $\mathcal{W}^D$ which is independent of droplet size), $\mathcal{D}_L$ always lands on the back side of $\mathcal{W}_S^D$ (relative to the direction of wave propagation) and experiences a propulsion $\textit{\textbf{p}}$ towards upper right (i). The interaction between $\mathcal{D}_S$ and $\mathcal{W}_L^D$ takes place at $t=T$, by then $\mathcal{W}_L^D$ has evolved for a time interval $T-\Delta{}t$. Therefore, the impact of $\mathcal{D}_S$ is made on the front side of $\mathcal{W}_L^D$, while $\mathcal{W}_L^D$ is returning to its zero position. Consequently, $\mathcal{D}_S$ is propelled towards upper right, the same direction as $\mathcal{D}_L$ (ii).

For \textit{l} = 1, due to the change of the interdroplet binding to the ${\bar{d}}_l=\left(l-{\epsilon{}}_l\right){\lambda{}}_D\sim{}0.8{\lambda{}}_D$
regime, the delay of the bouncing phase of $\mathcal{D}_L$ results in the impact of $\mathcal{D}_L$ being made at the front side of $\mathcal{W}_S^D$ (iii) and that of $\mathcal{D}_S$ at the back side of $\mathcal{W}_L^D$ (iv), respectively. Under the long-range binding regimes, both droplets feel a propulsion towards the upper left (Fig. 7c). It thus can be seen that the propulsions resulted from the interdroplet particle-wave association always propel the two droplets of the heterodimers to move towards the same horizontal direction. Moreover, the direction of the propulsion will be switched between the short-range binding regime and the long-range binding regimes due to the spatial change between the droplets.

\begin{figure*}
\centering
\includegraphics[width=0.8\linewidth]{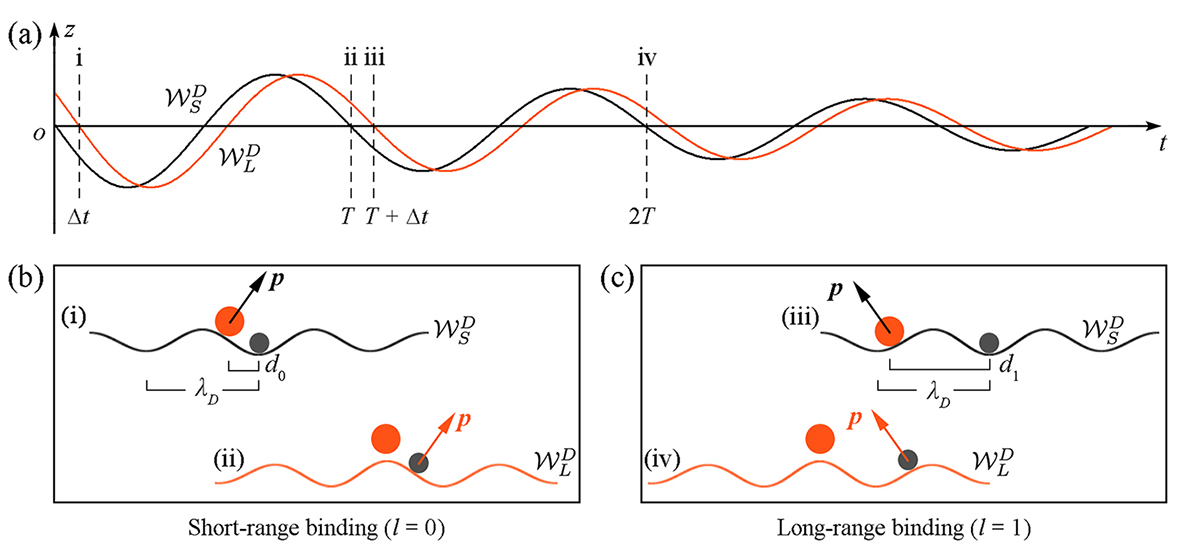}
\label{FIG7}
\caption{
(a) Schematics of the temporal phase shift between $\mathcal{W}^D$ induced by the impact of $\mathcal{D}_S$ (black) and $\mathcal{D}_L$ (red). (b), (c) Directional propulsion resulted from the interaction between individual droplets and the wave field generated by their partners in the \textit{l} = 0 regime (b) and the \textit{l} = 1 regime (c). The characteristic time of impact for cases i-iv is indicated in (a). The arrows indicate the direction of the propulsion that the local wave field exerts on its partner droplet during impact. Here (b) and (c) are in the one-dimensional spatial reference frame along the orbit.
}
\end{figure*}

\section{Linking horizontal orbiting and vertical bouncing}
\label{6_horizontal-vertical}
The quantization of the orbital motions now can be written as:

\begin{equation}
\label{eqdl2}
\bar{d}_{l}=\left\{\begin{array}{ll}{\left(d_{L}+d_{S}\right) / 2,} & {l=0} \\ {\left(l-\varepsilon_{l}\right) \lambda_{D},} & {l=1,2,3, \ldots}\end{array}\right.
\end{equation}

\begin{equation}
\label{eqrn2}
r_{n}=\left(n-\varepsilon_{n}\right) \lambda_{G}, \quad n=1,2,3, \ldots
\end{equation}
As discussed earlier, in Eq. (\ref{eqrn2}) a simplified linear correlation is used to approximate $\textit{J}_0$(2$\pi$$\textit{r}_n$/$\lambda{}_G$) = 0. Eq. (\ref{eqdl2}) and Eq. (\ref{eqrn2}) give clear indication that the quantization of the binding level and the orbit level results from the interaction of the bouncing droplets with two wave fields of different origins, which is a signature of the integrated pilot-wave droplet-bath system. The offsets ${\epsilon{}}_l=0.2$ and ${\epsilon{}}_n=0.4$ are thought to result from the self-tuning by the heterodimers in order to stabilize the two types of particle-wave associations \cite[]{protiere2006particle, eddi2009archimedean, filoux2015strings}. For instance, the association between the droplets and the wave emitted by their partners should self-adjust to maintain their collective motion when counteracting the locally changing current of the vortex pair along the orbits, which manifests as the observed instantaneous velocity and binding distance changes in Fig. 2(b) and Fig. 3(a). Moreover, the association between the heterodimers with the global wave field should also self-adapt to orbit radii that ensure dynamically stable correlation between their orbiting velocity and centripetal force in a given orbit level. It is of course reasonable to expect mutual influences between these self-tuning behaviors. Consequently, although the integrated pilot-wave field has a complex energy landscape, the heterodimers are able to maintain their well-defined yet dynamically-stable collective motions.

We proceed to investigate the relation between $\theta{}/2\pi{}$ and $\bar{v}$ as a function of the driving acceleration ${\gamma{}}_0$. Four most stable orbiting regimes (\textit{l}, \textit{n}), namely the (0, 2), (0, 3), (1, 2) and (1, 3) regimes are selected and compared. The driving acceleration is increased progressively from the bouncing threshold ${\gamma{}}_B\sim{}0.10$\textsl{\textrm{g}} to the Faraday threshold ${\gamma{}}_F\sim{}0.32$\textsl{\textrm{g}}, with a step size of 0.02\textsl{\textrm{g}}. To obtain $\theta{}$ and $\bar{v}$, the trajectories of the vertical bouncing motion and the horizontal orbiting motion of the heterodimers under each condition are captured by high-speed imaging and the images are analyzed using visual tracking. Nearly 500 heterodimers (1000 droplets) in total are analyzed, and the size distribution of the droplets is presented in Fig. 8(a) and Fig. 8(b). Due to the size filtering effect of the vibrating bath based on its effective acceleration \cite[]{gilet2007controlling}, which in the current case is determined by both ${\gamma{}}_0$ and \textit{n} (level splitting), the droplet diameter \textit{d} in our experiment is restricted to a narrow range from $\sim$1.0 mm to $\sim$1.6 mm. In addition, the mutual influence between the two droplets in the heterodimers further imposes restrictions on the droplet size selection. As a result, the size difference between $\mathcal{D}_L$ and $\mathcal{D}_S$ is found to be small, with the average of their diameter ratio \textit{$d_S/d_L$} close to 0.9 (inset of Fig. 8b).

\begin{figure*}
\centering
\includegraphics[width=0.82\linewidth]{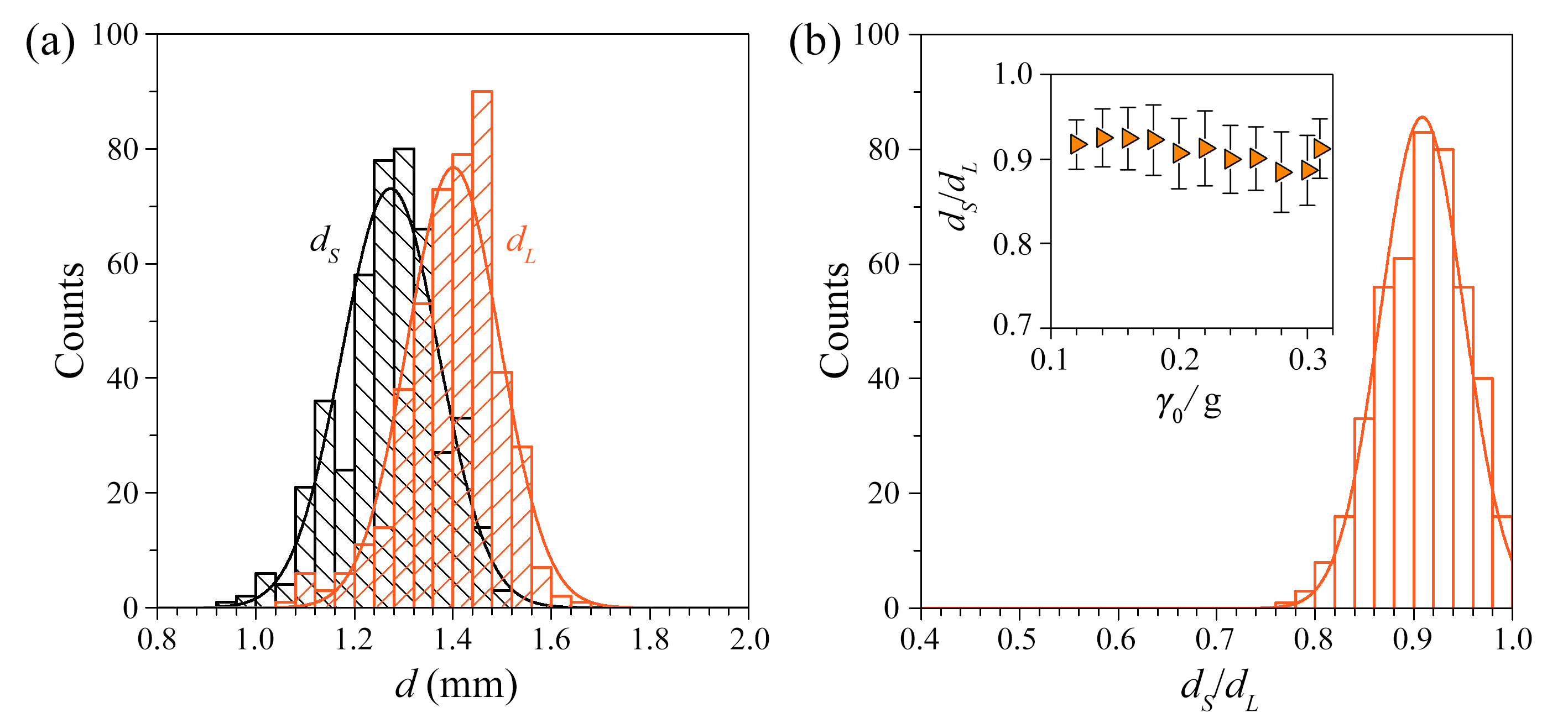}
\label{FIG8}
\caption{(a) A survey of droplet size distribution. (b) Relative droplet size and its acceleration dependence (inset).}
\end{figure*}

The delay of the impact of the large droplet to that of the small is found to hold for all the surveyed heterodimers and all the droplets are in the simple bouncing regime, even at high accelerations. As shown in Fig. 9(a) and Fig. 9(b), correlations are found between the maxima of $\theta{}/2\pi{}$ and ${\gamma{}}_0$/\textsl{\textrm{g}}, and the two types of binding regimes ($l=0$ and $l=1$) lead to different acceleration dependence. In the short-range binding regime ($l=0$), the phase shift maxima increase almost linearly with ${\gamma{}}_0$/\textsl{\textrm{g}} (Fig. 9a), while for the long-range binding regimes ($l=1$), the maxima show a parabola-like acceleration dependence (Fig. 9b). Note that multiple phase differences are possible for a given acceleration in each regime as can be seen from Fig. 9(a) and Fig. 9(b). This is a result of different droplet size of the heterodimers. The pentagrams in Fig. 9(b) further demonstrate the gradual decrease of $\theta{}/2\pi{}$ from about 0.1 at ${\gamma{}}_0$/\textsl{\textrm{g}} = 0.24 to near zero at ${\gamma{}}_0$/\textsl{\textrm{g}} = 0.30 for a (1, 3) regime heterodimer that we are able to maintain for longer than 1 h. When $\theta{}/2\pi{}$ becomes near zero, the heterodimer stops its orbital motion.

The influence of the global wave field $\mathcal{W}^G$ on $\theta{}/2\pi{}$ and $\bar{v}$ can be seen from the heterodimers in the same binding level \textit{l} but different orbit level \textit{n}, since the effective acceleration of $\mathcal{W}^G$ decreases as \textit{n} increases due to level splitting (Fig. 4c). Comparing the phase shift of the (0, 2) regime with the (0, 3) regime (Fig. 9a) or the (1, 2) regime with the (1, 3) regime (Fig. 9b), one finds that for the same binding regime, moving from one orbit level to another will not change the acceleration dependence of the phase shift. The main influence is a downshift of its maxima at higher orbit level \textit{n}. This is in good agreement with the energy landscape of the orbits of $\mathcal{W}^G$ given level splitting. As shown in Fig. 9(c) and Fig. 9(d), the surveyed average orbiting velocity $\bar{v}$ in different binding regimes shares very similar acceleration dependence with $\theta{}/2\pi{}$: The maxima of $\bar{v}$ also show a near-linear and a parabola-like acceleration dependence for the $l=0$ and $l=1$ regimes, respectively. Such similar acceleration dependence between $\theta{}/2\pi{}$ and $\bar{v}$ implies that the vertical bouncing of the heterodimers and their horizontal orbiting motion is directly related, which reflects the fact that the horizontal motion originates from the vertical bouncing of the droplets. The different particle-wave association regimes between the two types of inter-droplet binding regimes should be responsible for their different acceleration dependence. 

\begin{figure*}
\centering
\includegraphics[width=0.8\linewidth]{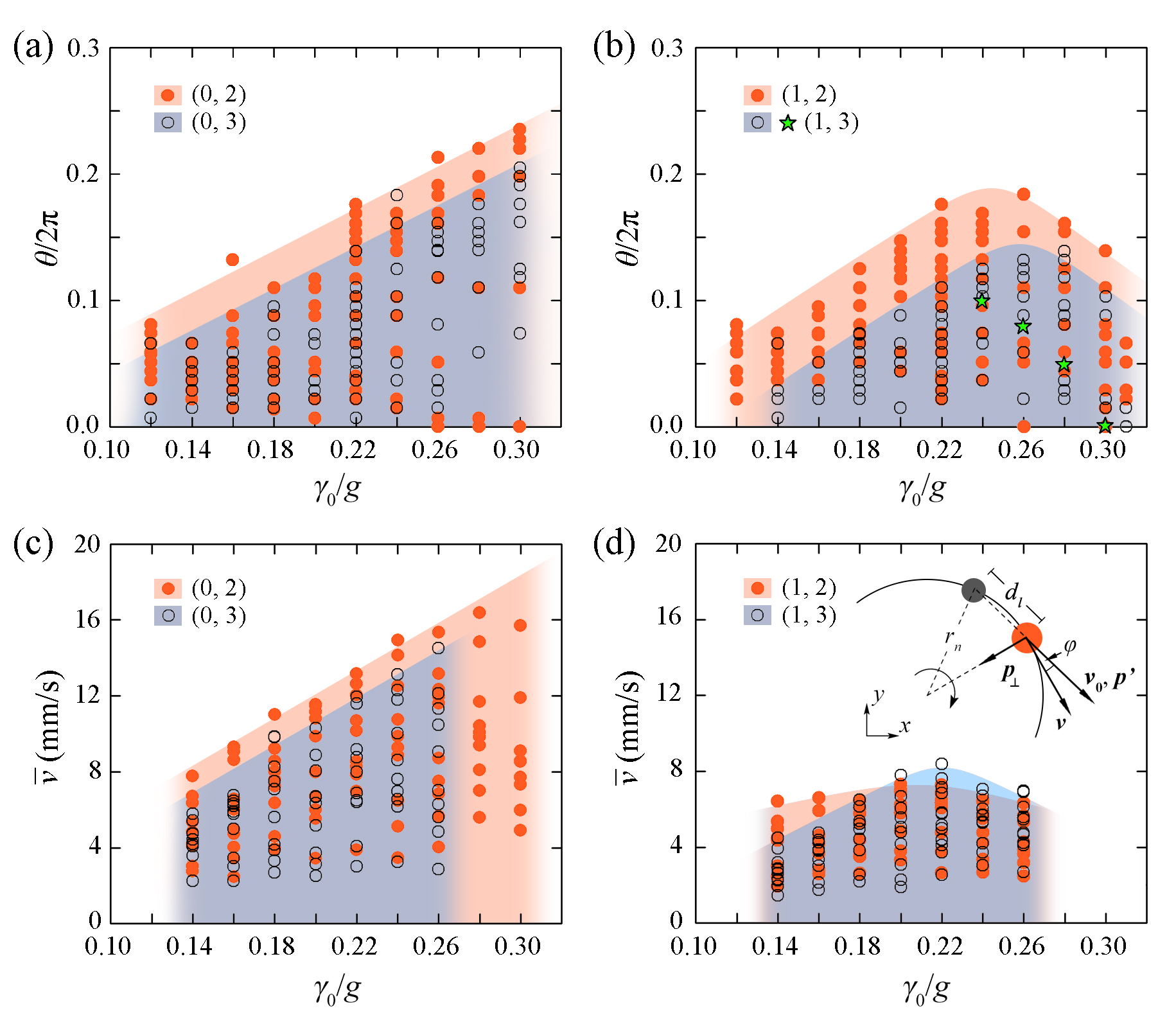}
\label{FIG9}
\caption{
(a), (b) Acceleration dependence of $\theta{}/2\pi{}$ for the (0, 2), (0, 3) regimes (a) and the (1, 2), (1, 3) regimes (b). The error in the measurement of $\theta{}/2\pi{}$ is 0.02. The pentagrams show the acceleration dependent phase shift of the same heterodimer. (c), (d) Acceleration dependence of $\bar{v}$ for the (0, 2), (0, 3) regimes (c), and the (1, 2), (1, 3) regimes (d). Error in $\bar{v}$ measurement $\textless$ 0.04 mm s$^{-1}$. The scatters represent the measured values, based on which the filled regions are drawn as guides for the eyes. The inset of (d) depicts the geometrical constraints which shows how the tangent projection of the droplet momentum is influenced by the binding distance $\textit{d}_l$ and orbit radius $\textit{r}_n$.
}
\end{figure*}

Noticeably, similar influence of \textit{n} on the $\bar{v}$ maxima is also found to hold for the $l=0$ regimes but breaks down for the $l=1$ regimes, that is, instead of showing an overall downshift of $\bar{v}$ maxima in response to their acceleration dependence in Fig. 9(b), the heterodimers in the (1, 3) regime demonstrate higher velocity than the (1, 2) regime under certain driving conditions (Fig. 9d). Such scenarios observed in the long-range binding regimes are not violations of our discussions and the observation can indeed be understood by further considering the enhanced spatial influences on droplet motion in the long-range binding regimes. In the orbiting plane, the direction of $\textit{\textbf{p'}}$ (the horizontal component of $\textit{\textbf{p}}$ resulting from interdroplet particle-wave association) is in the two-droplet-alignment, which is not the same as $\textit{\textbf{v}}$ (along the tangent of the orbit, see inset of Fig. 9d). This requires the momentum of the droplet to be projected partially to the tangent direction and partially to the normal direction of the orbiting plane (the $x-y$ plane). The normal projection is compensated for during the droplet-bath impact, which results in the centripetal force $\textit{\textbf{p}}_{\perp{}}$ needed for maintaining the orbital motion. Given the superposition of waves, we consider the influence of the global wave field $\mathcal{W}^G$ and the local wave field $\mathcal{W}^D$ separately. We also assume that both $\mathcal{W}^G$ and $\mathcal{W}^D$ induce similar $\bar{v}-{\gamma{}}_0$ correlations for both the short-range binding regime and the long-range binding regimes, through modulating the effective acceleration only (i.e., when the geometrical configurations are not considered). The geometrical configurations determine how the heterodimers distribute the tangent and normal projections, thereby affecting their orbiting velocity. The tangent projection of the velocity resulted from interdroplet particle-wave association scales as $v=v_0$cos${\varphi{}}_{\left(l, n\right)}$, where ${\varphi{}}_{\left(l, n\right)}$ is the angle between $\textit{\textbf{p'}}$ and $\textit{\textbf{v}}$. The geometrical constrains of the horizontal motion lead to
\begin{equation}
\label{eqphi}
\varphi_{(l, n)}=\arcsin \left(d_{l} / 2 r_{n}\right)
\end{equation}

Following Eq.(\ref{eqphi}), besides the effective acceleration which determines tangent velocity $v_0$, the tangent projection is also affected by both quantized regime parameters $d_l$ and $r_n$. Moreover, the two have opposite influences. The tangent projection is suppressed (increase in ${\varphi{}}_{\left(l, n\right)}$) by increasing the binding distance $d_l$ but favored by increasing the orbit radius $r_n$. As can be deduced based on Eqs.(\ref{eqdl2})$-$(\ref{eqphi}), the change between ${\varphi{}}_{\left(0,2\right)}\sim7.5^\circ{}$ and ${\varphi{}}_{\left(0,3\right)}\sim4.5^\circ{}$ is small when the orbit level moves from \textit{n} = 2 to \textit{n} = 3 in the short-range binding regimes. This is because in this case $d_0$ is small and $v$ is mainly determined by the \textit{n}$-$dependent effective acceleration. Therefore, the achievable velocity decreases as the orbit level moves from \textit{n} = 2 to \textit{n} = 3, due to a drop in the effective acceleration (Fig. 9c).

As $d_l$ significantly increases in the long-range binding regimes (Fig. 9d), the same orbit level change causes significant drop in ${\varphi{}}_{\left(l,\
n\right)}$, from ${\varphi{}}_{\left(1,\ 2\right)}\sim30.5^\circ{}$ to ${\varphi{}}_{\left(1,3\right)}\sim18.2^\circ{}$. In this case, the geometrical influence is magnified due to increased $d_l$, which becomes comparable to that of the effective acceleration. Since high orbit level \textit{n} favors the tangent projection, particularly for large $d_l$, the (1, 3) regime is able to achieve a higher velocity than the (1, 2) regime under certain driving conditions, which explains the different orbit-level dependence between Fig. 9(b) and Fig. 9(d). It can also be inferred from Eq.(\ref{eqphi}) that a more significant tangent velocity projection should be one of the reasons for the short-range binding regime to have a higher velocity than the long-range binding regimes. Another reason should be their more energetic particle-wave association (less damping). Both of them are caused by the change in the inter-droplet binding distance $d_l$. Similarly, the analyses can also be extended to higher-level long-range binding regimes ($l=2, 3, 4, ...$) by considering more significant influence of their regime parameters on both ${\varphi{}}_{(\textit{l},
\textit{n})}$ and damping.

\section{Discussions}
\label{7_discussions}
We have demonstrated the quantized orbital motion of dissimilar droplet pairs in the liquid metal droplet-bath system where the droplet motions are directed by a dual pilot-wave field. The quantization of the orbit radius and the binding distance is shown to be a result of the droplets interacting with the waves formed by meniscus oscillating and the impact of their partner droplets, respectively. The origin of the horizontal motion lies in a temporal bouncing phase shift between the two droplets of the chasing heterodimers due to their size mismatch. As shown by \textcite[]{molacek2013drops} and \textcite{couchman2019bouncing}, this is required by the vertical dynamics for the dissimilar droplets with different contact time and restitution coefficient to bounce in resonance with the vibrating bath. Our experiments evidence that the result of the mismatched bouncing is a delay of the impact of the large droplet to that of the small one, and, more importantly, the breaking of bouncing symmetry is a mechanism for initiating directional horizontal droplet motion. As discussed by \textcite[]{eddi2008wave, galeano2018ratcheting}, from the perspective of symmetry breakdown, the liquid metal heterodimers share similarity with the ratcheting droplet pairs. However, there are also major differences between the two types of motions in terms of directionality, reversibility, and binding regime. Note that in the current study we limit the vertical phase comparison to the bouncing phase shift between the two droplets in the heterodimers. In order to reach a better (more comprehensive) understanding of phase evolution in the system, it is worthwhile for future studies to further include the vibrating phase of the shaker (driving phase) as well as the bath wave field.

The interaction of the heterodimers with the two pilot-wave fields should both be considered to understand the observed quantized, directional, in-orbit chasing motions. The interaction between the droplets and the local wave field generated by their partners determines the chasing direction, through different particle-wave association regimes characterized by the interdroplet spatial binding. Different from the binding of walkers, the quantization of interdroplet binding distance in our experiment is found to only take successive integers of
${\lambda{}}_D$ (no half-wavelength binding regime has been observed). The relatively small achievable bouncing phase shift of the chasing liquid metal heterodimers, $\theta{}/2\pi{}<1/4$ throughout the applied acceleration range (Figs. 9a, 9b), might be responsible for such a difference, since half-wavelength binding typically occurs when the droplets are bouncing in antiphase ($\theta{}/2\pi{}=1/2$) \cite[]{protiere2006particle, couder2005dynamical, protiere2008exotic, eddi2009archimedean, eddi2012level, filoux2015strings}. The vertical trajectories of the droplets show that their bouncing is synchronized with the bath vibration as well as the meniscus oscillating, featuring the driving frequency \textit{f}. This indicates that both the waves generated by the meniscus oscillating and by the droplet bouncing are harmonic. Furthermore, the droplets in the current system mainly adopt the simple bouncing mode, which coincides with our previous report \cite[]{zhao2018electrically}.  Due to the dominance of the single bouncing regime, the chasing direction of the heterodimers shows no dependence to applied acceleration.There are three reasons that could possibly account for the absence of the walking and other bouncing regimes in the current system: (1) The peculiar fluid properties of the liquid metal in comparison to the commonly used silicon oils; (2) Instead of a gaseous environment (air), the liquid metal droplets bounce on a liquid-liquid interface and their free flight takes place in a liquid phase, and, consequently, the droplets are expected to experience distinct drag and lubricating force due to significant differences in viscosity and compressibility of the surrounding medium; and (3) Influenced by the meniscus-induced waves, the bath becomes unstable before the transition states being reached.

The interaction between the droplets and the annular global wave field provides the confinement to lock the horizontal motion of the heterodimers into circular orbits, which would otherwise be in an arbitrary direction. We note that, although kept unchanged in the current experiments, ${\lambda{}}_G$ can be readily adjusted through frequency control. Allow for the superposition nature of waves and their tunability, guiding droplets with an integrated pilot-wave field could be a promising strategy for future works to reveal the unexploited potentials of the droplet-bath system. In this regard, our findings could provide helpful guidelines for advancing the pilot-wave hydrodynamics and extending its implications to other physical systems. The striking similarities between the heterodimers in the current hydrodynamic system and the optical system are convincing as to the driving role of the local wave field and the confining role of the global wave field for particle/droplet guiding in both systems. The evidence unveils symmetry breakdown as a universal mechanism for motion initiating in wave-mediated systems.

\begin{acknowledgments}
\label{acknowledgments}
This work was supported by the NSFC Key Project under Grant No. 91748206, Dean’s Research Funding and the Frontier Project of the Chinese Academy of Sciences. The authors are in debt to Dr Junyu Gao  (Institute of Automation, Chinese Academy of Sciences) for his helpful advice on droplet tracking and image analysis.
\end{acknowledgments}

\bibliography{Tang_et_al_manuscript}

\end{document}